\documentclass[preprint]{aastex}
\shorttitle{Jet-CH collision and associated CME}
\slugcomment{Submitted to ApJL}
\shortauthors{Zheng et al.}
\begin{document}

%\title{{\bf Jet-coronal hole collision and a closely-related coronal mass ejection}}
\title{Solar jet-coronal hole collision and a related coronal mass ejection}
\author{Ruisheng Zheng, Yao Chen, Guohui Du, and Chuanyang Li}
\affil{Shandong Provincial Key Laboratory of Optical Astronomy and
Solar-Terrestrial Environment, and Institute of Space Sciences,
Shandong University, 264209, Weihai, China; ruishengzheng@sdu.edu.cn}

\begin{abstract}
Jets are defined as impulsive, well-collimated upflows, occurring
in different layers of the solar atmosphere with different scales.
Their relationship with coronal mass ejections (CMEs), another
type of solar impulsive events, remains elusive. Using the
high-quality imaging data of AIA/SDO, here we show a well-observed
coronal jet event, in which part of the jets, with the embedding
coronal loops, runs into a nearby coronal hole (CH) and gets
bounced towards the opposite direction. This is evidenced by the
flat-shape of the jet front during its interaction with the CH and
the V-shaped feature in the time-slice plot of the interaction
region. About a half-hour later, a CME initially with a narrow and jet-like 
front is observed by the LASCO C2 coronagraph, propagating along
the direction of the post-collision jet. 
We also observe some 304~{\AA} dark material flowing
from the jet-CH interaction region towards the CME. We thus
suggest that the jet and the CME are
physically connected, with the jet-CH collision and the
large-scale magnetic topology of the CH being important to define
the eventual propagating direction of this particular jet-CME
eruption.
\end{abstract}

\keywords{Sun: activity --- Sun: corona --- Sun: coronal mass ejections (CMEs)}

\section{Introduction}
Solar coronal jets, first observed in X-rays with the Soft X-ray
Telescope (XRT, Tsuneta et al. 1991) on board the Yohkoh
satellite, represent a group of impulsive events characterized by
well-collimated upflows with different scales developing in
different layers of the solar atmosphere (e.g., Shibata et al.
1992; Savcheva et al. 2007; Chen et al. 2012). They are generally
believed to be energized by magnetic reconnection, often
associated with an inverse Y-shaped, anemone-like configuration
involving open field lines in coronal holes (CHs) or open-like
large-scale closed loops extending from an active region (Shibata
et al. 1992, 2007; Schmieder et al. 1995; Rachmeler et al. 2010;
Pariat et al. 2009, 2015). These open or open-like field lines are
important in collimating jets.

Coronal mass ejections (CMEs) are another type of impulsive energy
release events in the solar atmosphere, with a much larger scale
and stronger impact on nearby coronal structures such as streamers
and CHs. There exist a number of studies examining the strong CME
disturbance to coronal streamers (e.g., Hundhausen et al. 1987;
Sheeley et al. 2000; Tripathi \& Raouafi 2007; Chen et al. 2010).
In the meantime, both streamers and CHs have been suggested to
have effects on the propagating direction of CMEs, a crucial
factor determining the CME geo-effectiveness. For instance,
Gopalswamy et al. (2009) reported events with sources very close
to the solar disk center that are unexpectedly \emph{NOT}
associated with interplanetary CMEs (yet accompanied by
interplanetary shocks), and they attributed this to possible
interaction and further deflection of CMEs by nearby CH(s).
Nevertheless, a direct observation of this CME-CH interaction
process remains absent. Neither do we know how and where the
deflection takes place.

While both jet and CME represent impulsive ejection of plasmas to
upper levels of the solar atmosphere, their relationship remains
obscure. Can a relatively large-and-fast jet drive a CME or can
the CME actually trigger some jets, by, for example, opening
initially-closed magnetic field? Different scenarios have been
developed (e.g., Pariat et al. 2009, 2015), and actual answers may
differ to different event, depending on specific
circumstances{\footnote{Note that during the peer-review process
of our manuscript, a study reporting a CME event likely triggered
by a coronal jet was published (Liu et al., 2015), which presents
evidence supporting the close relation between a jet and a CME,
with the jet pushing some overlying blob-like magnetic structure
which later becomes the CME front and the jet likely evolves into
the CME core.}}. It is also very interesting to ask, considering
the above-mentioned possibility of strong CME-CH interaction, can
a jet, if moving along large-scale active region loops, actually
interact with a nearby CH? None of this kind of event has been
reported ever. In this study, we present unambiguous evidence of
such an event, revealing the collision between a set of coronal
jets and a nearby CH with high quality data from the Atmospheric
Imaging Assembly (AIA: Lemen et al. 2012) onboard the Solar
Dynamics Observatory (SDO: Pesnell et al. 2012). It turns out that
part of the jets is reflected towards the opposite direction, and
this dynamical jet-CH interaction may have led to a successful
eruption along the same direction.

\section{Observations and Data Analysis}
We mainly analyzed the AIA/SDO data that provides the essential
observations of the event. The AIA instrument has ten EUV and UV
wavelengths, covering a wide range of temperatures. The AIA
observes the full disk (4096~$\times$ 4096 pixels) of the Sun and
up to 0.5 $R_\odot$ above the limb, with a pixel resolution of
0.6$"$ and a cadence of 12 s. The eruption is visible in all AIA
EUV channels. The passbands of interest here are 131~{\AA} (Fe XXI, $\sim$10 MK), 211~{\AA} (Fe XIV, $\sim$2.0 MK), 171~{\AA}
(Fe IX, $\sim$0.6 MK), and 304~{\AA} (He II, $\sim$0.05 MK).
Magnetograms and intensity maps from the Helioseismic and Magnetic
Imager (HMI: Scherrer et al. 2012), with a cadence of 45 s and pixel scale of
0.6$"$, were used to check the magnetic field configuration of the
source region. The CME evolution in the high corona was captured
by the Large Angle and Spectrometric Coronagraph (LASCO) C2
(Brueckner et al. 1995).

The kinematics of the jets and associated mass flow were analyzed
with the time-slice approach. The speeds were determined by linear
fits, with error bars given by the measurement uncertainty that is
assumed to be 4 pixel ($\sim1.74$ Mm) for AIA data. We also used
the Potential Field Source Surface (PFSS: Schrijver {\&} De Rosa
2003) model to extrapolate the HMI photospheric field measurement
to describe the large-scale magnetic field geometry.

\section{Results}
\subsection{Coronal jets}
The event occurred at the eastern boundary of the NOAA Active
Region (AR) 12403 on 25 August 2015. Upper panels of Fig.1 show
the AR image observed at AIA 211~{\AA}, and the intensity map and
magnetogram of HMI. The small white boxes ($\sim$S14E13) present
the source area in which jets originated. It can be seen that the
AR consists of a positive-polarity leading sunspot and a
negative-polarity following sunspot. In the region given by white
boxes, there exists a small parasitic positive polarity. An
elongated low-latitude elephant-trunk CH exists eastward of the AR
(white arrows in Fig.1a).

In Fig.1d-h, we present the sequence of HMI magnetogram from 03:30
UT to 11:30 UT to examine the magnetic evolution of the jet source
region. As a result of earlier magnetic flux emergence, a small
positive patch was embraced by negative dominant polarities and
became the parasitic polarity. Comparing these magnetograms, we
see that significant flux cancellation took place. This is further
confirmed by the temporal changes of positive and negative
magnetic fluxes in the FOV of Fig.1d-h, as plotted in Fig.1i.
Before 07:12 UT (dashed vertical line), the negative flux
increased continuously, while the positive counterpart did not
change much. After 07:12 UT, both fluxes started to decrease with
the positive one changing at a much steeper gradient. At about
10:20 UT (dotted vertical line), the positive flux presented an
even faster declining rate.

The later time was consistent with the onset of the jet event. In
the meantime, the GOES soft X-ray (SXR) profiles started to
increase after 10:23 UT. Two SXR peaks were recorded in the
following hour, corresponding to a C1.7 flare (peaking at 10:34
UT) and a C2.2 flare (peaking at 10:44 UT), respectively. Both
flares were associated with jet activities. This is consistent
with the general picture that jets are energized by magnetic
reconnection, as evidenced here by significant flux cancellation
and flare occurrence (e.g., Wang et al. 1998; Chae et al. 1999;
Chifor et al. 2008; Yang et al. 2011; Liu et al. 2011; Pariat et
al. 2009, 2015).

In Fig.2, we present the dynamical evolution of coronal jets
observed at AIA 171~{\AA}. As seen from Fig.2a-c and the
accompanying animation, the jet started from the southern end of
the bright flaring loops, exhibiting a gradual footpoint migration
towards the northern end. The migration indicates an apparent
motion of the main flaring reconnection site. For the convenience
of description, we separate the jets into three subsequent
episodes, consisting of the initial relatively weak part (J1,
starting at 10:26 UT), the middle part which is the strongest one
and of particular interest to this study (J2, starting at 10:30
UT), and the third part which is basically confined by underlying
loops (J3, starting at 10:38 UT). Note that similar confining mass
flows have been used to trace the twisted structure internal of a
flux rope (e.g., Li \& Zhang 2013; Yang et al. 2014). These
episodes of jets have been pointed out in Fig.2a-c. As mentioned,
the jets mainly emanated from the FOV of Fig.1d-h, above the
parasitic polarity. This is a general source property prescribed
in jet modelling (e.g., Pariat et al. 2009, 2015). Because J1 was
relatively weak and J3 was mostly confined, here we focus our
study on J2.

Using the time-slice approach along the dotted line (S1) in Fig.2c, 
the derived velocity of J2 is close to 500 km s$^{-1}$
(Fig.2f), much faster than the statistical average speed of
$\sim$200 km s$^{-1}$ for jets (Shimojo et al. 1996). It is clear
that J2 lasted for $\sim$ 30 mins with continuous mass ejection.
J2 initially moved along its associated AR loops, and carried the
loops to extend. It is interesting to see that the forward
extension of the jet-loop structure was suddenly stopped. The
curved side of the jet-loop structure became flat-shaped with
kinks at both ends, and part of J2 was clearly bounced towards the
opposite direction while the left part returned to the solar
surface (best seen in the online animation). The first sign of
bounced-back material was present around 10:38 UT as seen from the
online animation. The flat-shaped feature appeared around the
interface between the nearby CH and the east edge of the AR,
indicating that the jet carrying the loop ran into the AR-CH
boundary and got reflected there. The reflection is also seen from
the height-time plot along slices S2 and S3 (short lines in
Fig.2d-e), from which we see a distinct V-type structure
(Fig.2g-h). The speeds of the jet-loop structure along S2 and S3
before and after the reflection are nearly the same ($\sim$90 km
s$^{-1}$).

Note that, after the jet-CH interaction, part of the jet material
was stagnated (black arrows in Fig.2g-h) while the left part
presented a signature of continuation of mass flows towards west
(white arrows in Fig.2d-e, see also the online animation), at a
fast speed of $\sim$400 km s$^{-1}$. See Fig.2i for the time-slice
plot along S4. This is only slightly slower than the pre-collision
jet. Yet, the jet front faded away shortly. So, it is not
known, at this time, whether the reflected jet flows have escaped
the corona or not.

%\subsection{CME and its relation with the jet}
\subsection{CME and its relation with the jet}
A weak CME feature appeared in the LASCO C2 FOV at 11:14 UT, with
a hardly identifiable narrow front, developing into a much clearer
CME structure in 10-20 mins (see Fig.3 and the accompanying
animation). The central position angle of the ejecta at 11:26 UT
was 238$^\circ$ (the angle increases counterclockwise with
0$^\circ$ along north). Initially, the CME presented a narrow
jet-like morphology and later became diffusive without a clear
flux-rope signature. It is thus difficult to determine the exact
type of this eruption (see Vourlidas et al. 2013). The appearance
time of a clear CME signature in C2 is about 50 mins later than
the first sign of the jet-CH collision ($\sim$ 10:38 UT).
In addition, the continuation part of the reflected jet
flow is basically toward the CME direction. This close
temporal-spatial correlation suggests that the jet may be
associated with the CME.

It is crucial to further figure out whether the jet front
continued its westward motion toward the solar limb to become a
part of the CME or it actually moved downwards along a curved loop
path and confined there. For short, we refer these suggestions to
be the eruptive picture and the confining picture, respectively.
In the following we present observational facts that, from our
point of view, favor the first possibility.

Firstly, from Fig.4 and accompanying animations, in which
a set of AIA images at 171, 211, and 131~{\AA} are presented, we
see that after 10:55 UT the jet front seemed to have moved beyond
the associated AR loop system before its eventual fading-away, as
pointed out by the arrows in Fig.4a-e, rather than returned to the
solar surface along a curved loop-like path if being confined.

Secondly, following the jet front, since 11:06 UT a
systematic westward motion of a set of loops started to appear, as
pointed by arrows in Fig.4f-h, with a speed of 10-20 km s$^{-1}$.
This is best seen from the online animations. The motion lasted
for more than 30 mins, possibly an effect of continuous stretching
exerted by the westward mass motion. Again, this is not
inconsistent with the eruptive picture.

Thirdly, from the 304~{\AA} data we observed an obvious outflow
of filament-like dark material after 11:20 UT (see the white box
in Fig.5a-c). It seems that the material corresponded to part of
the reflected jet, being stagnated and accumulated around the
jet-CH interaction region (best seen from the accompanying
animation). They became dark possibly due to a cooling process.
Note that there was no filament eruption observed during the
event, so these dark material was not due to any filament
eruption. Its outflow speed ($\sim$239 km s$^{-1}$) can be derived
using the distance-time analysis along the slice S5 (Fig.5d). The
304~{\AA} material moved out of the AR, along the direction
pointing to the CME. This traces the open path from the jet-CH
collision region to the CME, providing additional support to the
first picture. Note that due to the time delay, the 304~{\AA}
material could not become the CME front yet they may provide some
mass supply to the eruption.

The last observational fact worthy of mention is that no
any other detectable eruptive activities were present on the solar
disk according to all passband data of AIA, and on the back side
according to the Extreme Ultraviolet Imager (EUVI) (Howard et al.
2008) on board the twin spacecraft of Solar-Terrestrial Relations
Observatory (STEREO: Kaiser et al. 2008) with separation angles
from the Earth $172.348^\circ$ (STEREO-A) and $175.495^\circ$
(STEREO-B) at the time.

In summary, the above observational facts favor the first picture,
i.e., the post-collision jet further evolves into a part of the
CME. The distance from the jet-CH interaction region to the CME
front is $\sim$ 3 R$_\odot$, indicating an average projected
propagation speed of $\sim$ 700 km s$^{-1}$ if assuming the
post-collision jet front later becoming the CME front. This is
faster than the AIA-measured projection speed of the jet,
suggesting that either the jet gets further accelerated during its
outward propagation or the jet is not the counterpart of the CME
front and there exists other or earlier eruptive magnetic
structures ahead of the jet. It should be pointed out that how the
jet evolves into the CME and exactly which part of the CME
corresponds to the jet front remains not resolved with available
data set, partly due to the absence of CME signatures in the AIA
FOV.

Further examining the PFSS results of the CH-AR magnetic field
lines (Fig.3 c-d), we see that the CH open field lines are of
negative polarity and lying next to the closed loop system that is
rooted at the large negative polarity of the AR. The CH field
lines, with a strong non-radial expansion, occupy the space above
the closed AR loops. This magnetic configuration helps understand
how the observed initially-collimated jet (along the eastern edge
of the AR) runs into field lines of the nearby CH and then flows
outward along the specific trajectory.

\section{Summary}

Here we present a first-of-its-kind observational study on a
jet-CH colliding process showing that the post-collision jet was
reflected towards the opposite direction. We also present
compelling evidence supporting that the jet activity may have
developed into a successful eruption (i.e., a CME). The jet-CH
collision is evidenced by the flat morphology of the jet front
observed by AIA, while the jet-CME relation is supported by their
close temporal-spatial correlation, the observed outflow at
304~{\AA}, the large-scale CH-AR magnetic field configuration
given by PFSS, and the fact that no other identifiable eruptive
activities on the solar surface including the backside, among
other observations. The presumed jet-CME route basically
follows the over-expanding trend of the CH open field lines above
the AR according to PFSS extrapolation, indicating a strong role
played by the CH structure in defining the CME propagating
direction. This is consistent with earlier studies, which were
however not based on direct observation of CME-CH interaction,
that CHs are important in affecting the CME propagating direction
and thus the consequent geo-effectiveness. The study is possible
because of the unprecedented high-quality data of AIA/SDO.

\acknowledgments

SDO is a mission of NASA's Living With a Star Program. The authors
thank the SDO team for providing the data. This work is
supported by grants NSBRSF 2012CB825601, NNSFC 41274175, and
41331068, and Yunnan Province Natural Science Foundation 2013FB085.

\clearpage

\begin{figure}
\epsscale{0.9} \plotone{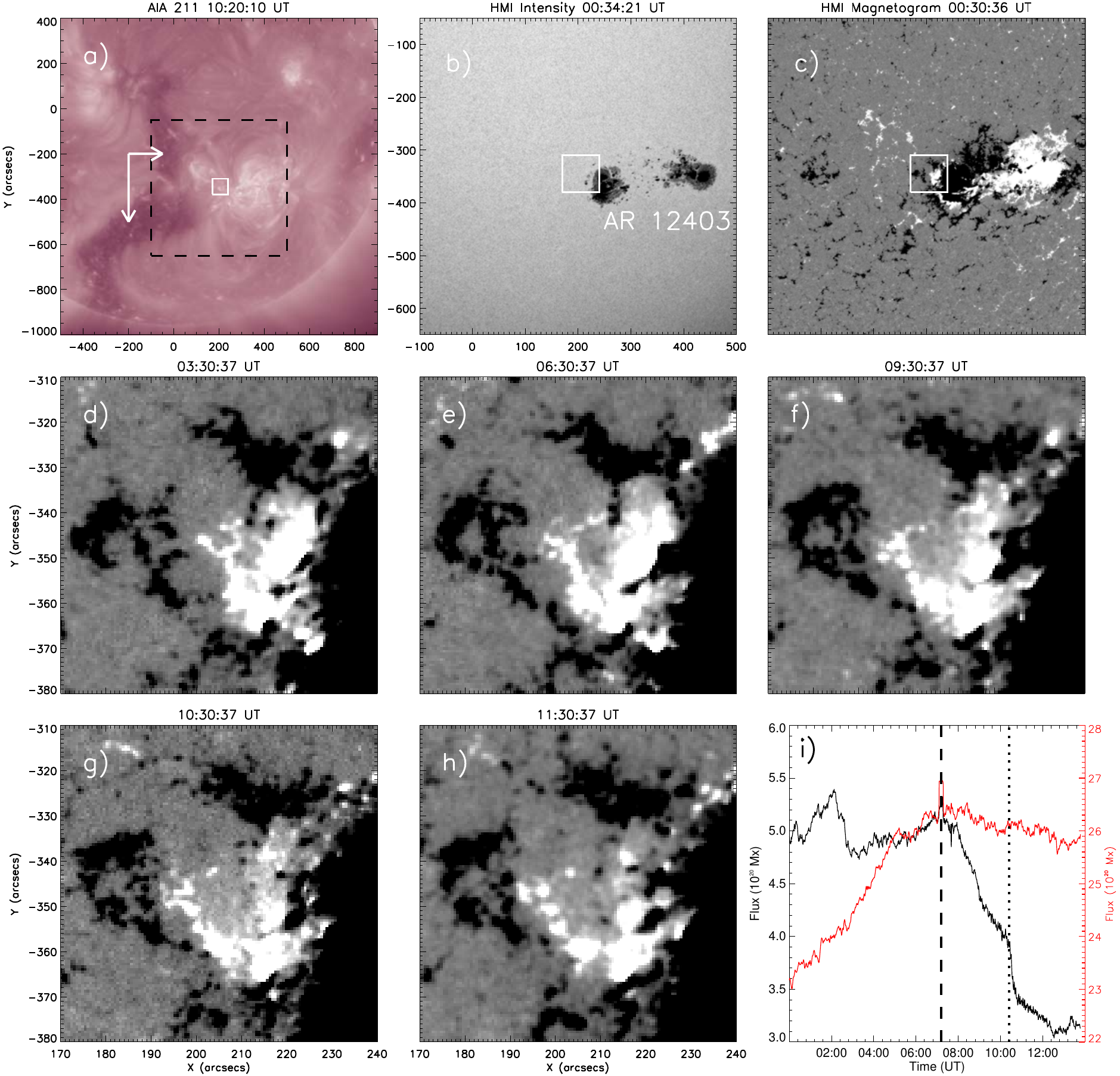} \caption{Overview of the NOAA AR
12043 and the magnetic evolution of the event. (a) AIA 211~{\AA}
image showing the eruption region (white box) and the adjacent
``elephant trunk" CH (white arrows), in which the black box
indicates the FOV of (b-c). (b-c) HMI intensity map and HMI
magnetogram showing the eruption region (white box) and the AR.
(d-h) A sequence of HMI magnetograms showing the magnetic
cancellation regions, in a FOV indicated by the box in panel c.
(i) The changes of the negative (red) and positive (black)
magnetic fluxes within the FOV of panels (d-h). The vertical
dashed and dotted lines mark the onset of the abrupt decreasing
and the beginning of the eruption, respectively. A color figure is
available online.\label{f1}}
\end{figure}

\clearpage

\begin{figure}
\epsscale{0.9} \plotone{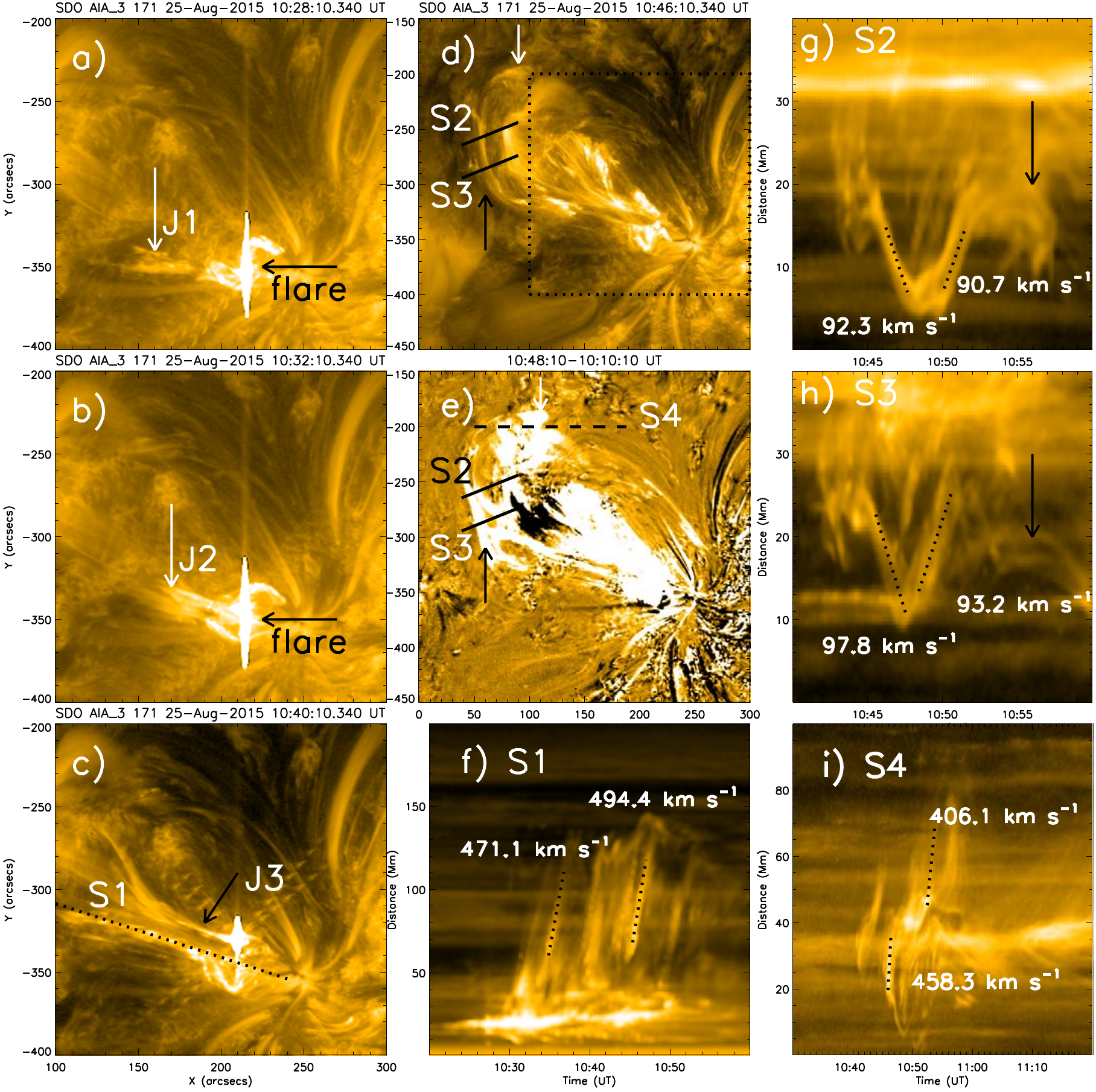} \caption{Coronal jets and their
interaction with the nearby CH, observed at AIA 171~{\AA}. (a-c)
Images showing the coronal jets (J1-J3) and the associated flaring
loops. (d-e) Images showing the jet-CH interaction. The short
lines (S2-S3) are used to construct slice-time plots revealing the
deflection of the jet-loop structure. The black dashed box in (d)
indicates the FOV of (a-c). (f) Slice-time plot showing the motion
of the J2 along S1. (g-h) Slice-time plots along slices S2-S3. The
flow stagnation is pointed out by black arrows. (i) Slice-time
plot along S4. The dotted lines are used to derive the
linearly-fitted speeds. A color figure and an accompanying
animation are available online.\label{f2}}
\end{figure}

\clearpage

\begin{figure}
\epsscale{0.9} \plotone{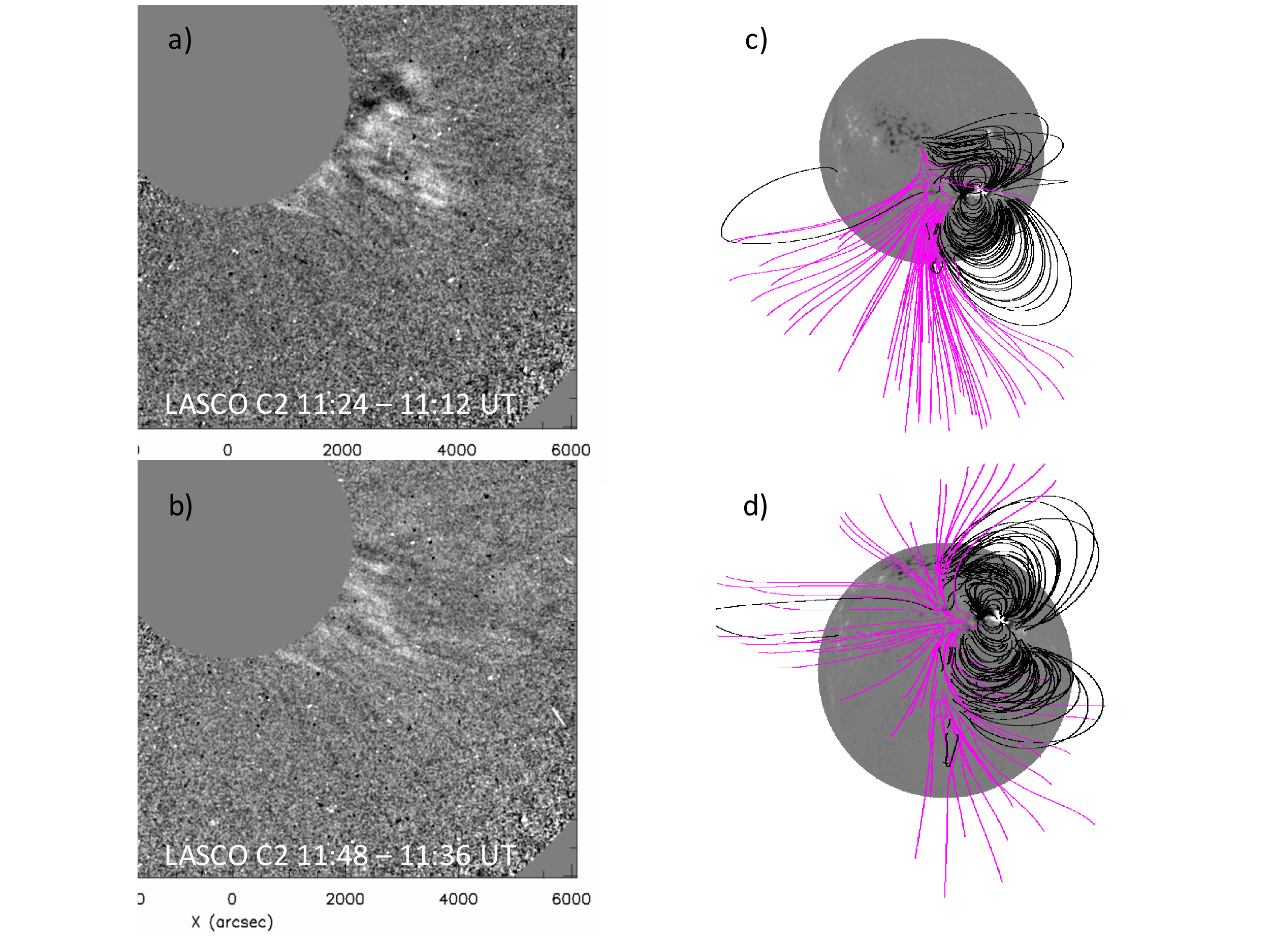} \caption{CME evolution and PFSS
extrapolation results. (a-b) LASCO C2 images of the CME. (c-d)
PFSS extrapolated field lines for the CH open field lines (purple)
and the active region closed field lines (black). Panel (c)
corresponds to the AIA FOV, while panel (d) is given by an upward
(northward) rotation of panel (c) by 50$^\circ$, to show up more
details of the over expansion of the CH magnetic field lines. A
color figure and an accompanying animation are available online.
\label{f3}}
\end{figure}

\clearpage

\begin{figure}
\epsscale{0.9} \plotone{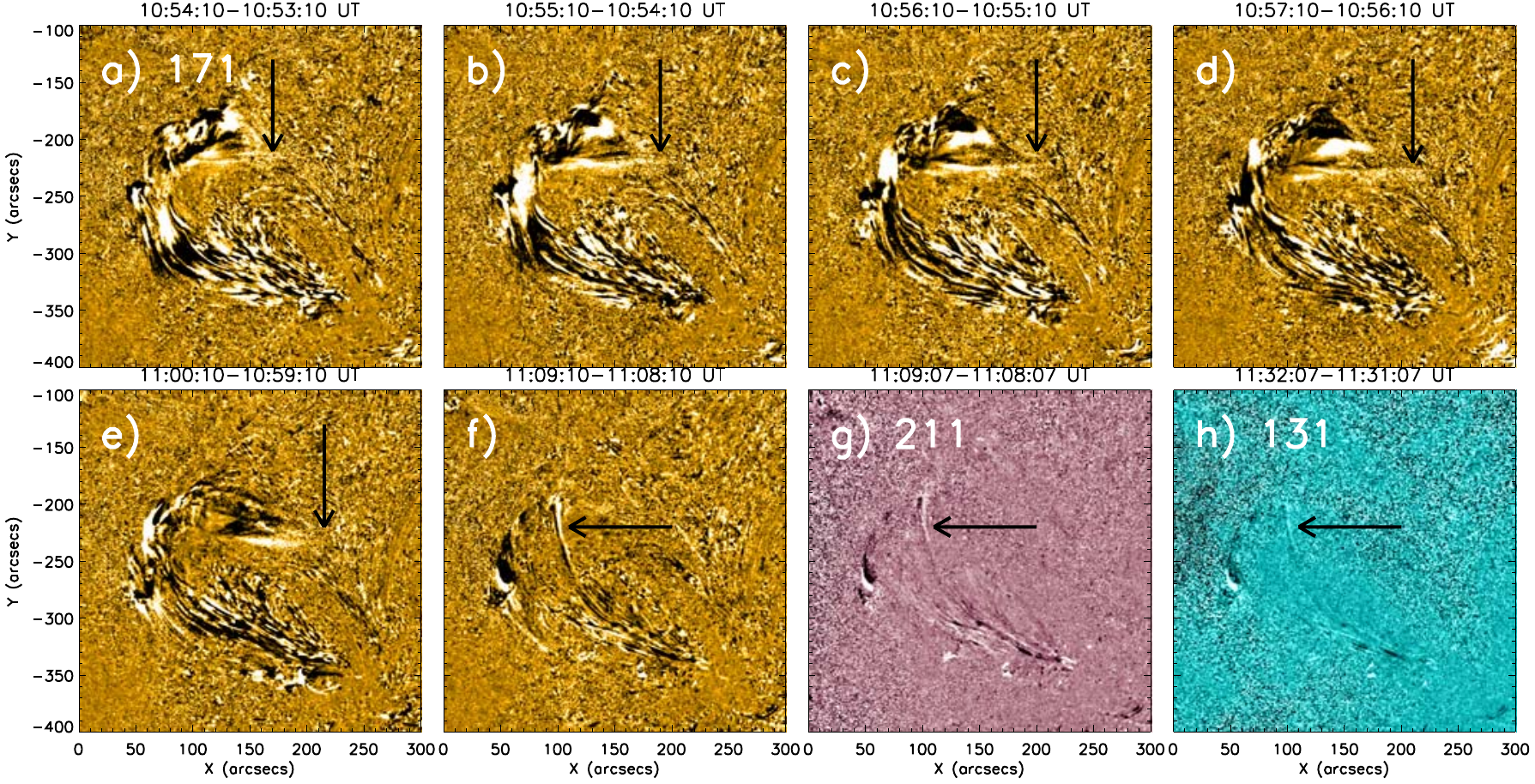} \caption{AIA images at 171
(a-d), 211 (e-f), and 131~{\AA} (g-h) highlighting the
continuation of the jet mass flow towards west and the associated
post-jet motion. Arrows in (a-d) point to the fast westward-moving
jet fronts, and arrows in (e-h) point to the relatively slow
westward-moving loops. A color figure and 3 accompanying
animations are available online. \label{f4}}
\end{figure}

\begin{figure}
\epsscale{0.9} \plotone{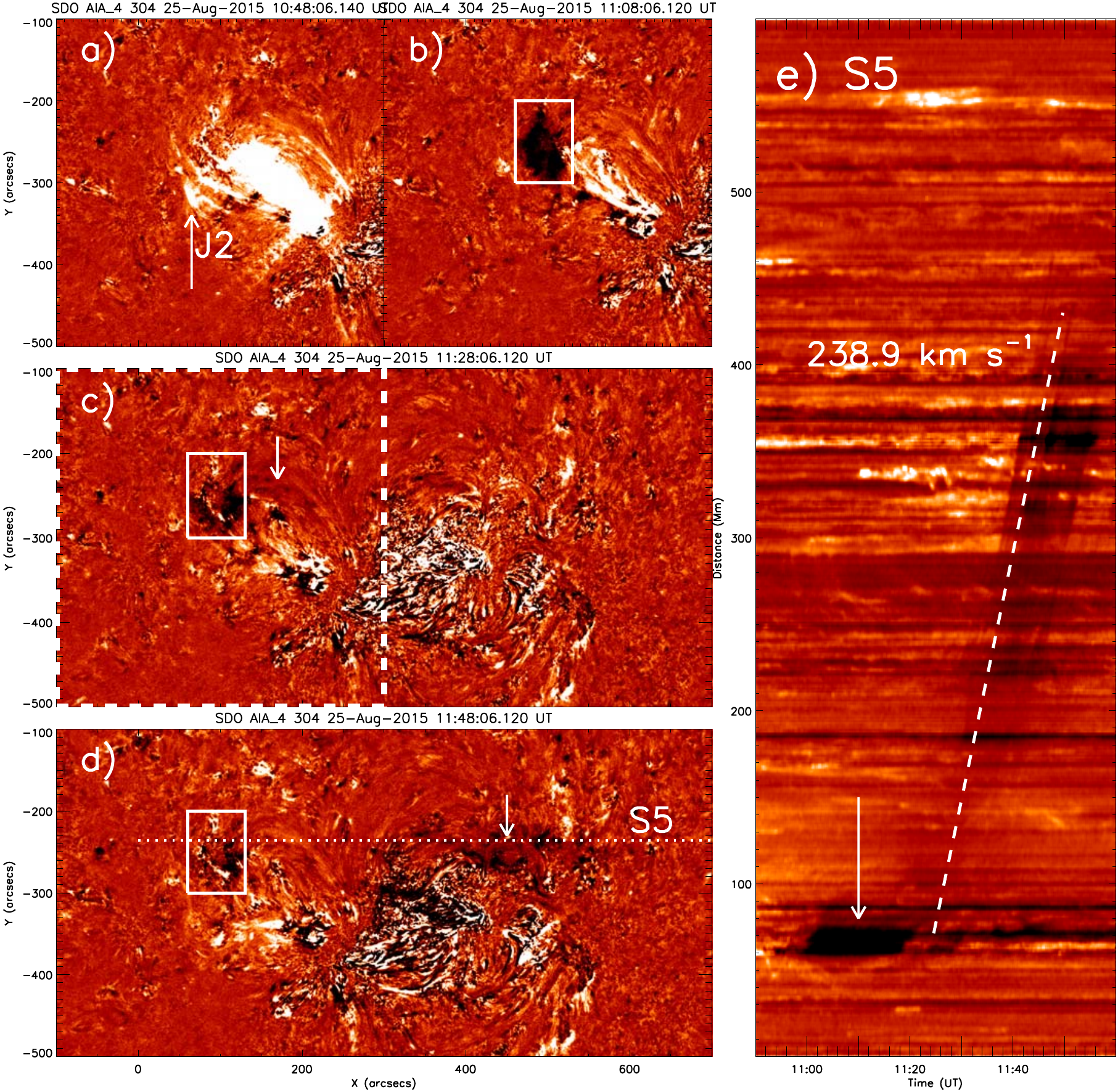} \caption{Accumulation and drift
of the dark filament-like mass observed at AIA 304~{\AA}. (a-b)
Base-difference images exhibiting the accumulation of the jet mass
as indicated by the white box. (c-d) Base-difference images
illustrating the drift of the 304~{\AA} material (white arrows).
The dotted line in panel (d) is selected to construct a slice-time
plot displaying the mass flow. The dashed box in panel (c)
indicates the FOV of panels (a-b). (e) Slice-time plot along the
slice S5. The dashed line is used to obtain the speed of mass
flow. A color figure and an accompanying animation are available
online. \label{f5}}
\end{figure}

\end{document}